\documentclass[aps,prb,twocolumn,showpacs,superscriptaddress]{revtex4}
\usepackage{graphicx,amssymb,amsmath}
\begin{document}
\title{
Photonic analog of graphene model and its extension \\
-- Dirac cone, symmetry, and edge states --
}
\author{Tetsuyuki Ochiai}
\affiliation{Quantum Dot Research Center, National Institute for
Materials Science (NIMS), Tsukuba 305-0044, Japan}
\author{Masaru Onoda}
\affiliation{Department of Electrical and Electronic Engineering,
Faculty of Engineering and Resource Science, Akita University, Akita 010-8502, Japan}
\date{\today}
\begin{abstract}
This paper presents a theoretical analysis on bulk and edge
 states in honeycomb lattice photonic
 crystals with and without time-reversal and/or space-inversion
 symmetries. 
Multiple Dirac cones are found in the photonic band structure and the
 mass gaps are  controllable via symmetry breaking.   
The zigzag and armchair edges of the photonic crystals can support 
novel edge states that reflect the
 symmetries of the photonic crystals. The dispersion relation and the
 field configuration of the edge states are analyzed in detail in
 comparison to electronic edge states. Leakage of the edge states to free
 space is inherent in photonic systems and is fully taken into account in the
 analysis. 
A topological relation between
 bulk and edge, which is analogous to that found in quantum Hall systems, 
is also verified.

\end{abstract}
\maketitle

\section{Introduction}
A mono-layer of graphite sheet, called graphene, has attracted growing
interests recently.\cite{novoselov2005tdg,geim2007rg}  Graphene exhibits a
Dirac cone with a linear dispersion 
at the corner of the first Brillouin zone, resulting in a
variety of novel transport phenomena of electrons.   They stimulate 
theoretical and experimental studies taking account of analogy to
physics of relativistic electron, such as  Klein tunneling\cite{klein} 
and Zitterbewegung\cite{Schrodinger}.  
Moreover, semi-infinite graphene and  finite stripe of 
graphene with zigzag edges support a novel edge state with nearly flat
dispersion.\cite{nakada1991esg} 
On the contrary, armchair edge does not support such an edge state.   
The flat dispersion implies that the density of state (DOS) diverges at the 
flat band energy, in a striking contrast to the zero DOS in bulk. 
So far, theoretical investigation of graphene heavily relies on the 
tight-binding model.  
It is not clear to what extent the edge states change in other models on 
the honeycomb lattice.

Here, we study a photonic analog of graphene model,\cite{cassagne1995pbg}
namely, two-dimensional photonic crystal (PhC) composed of 
the honeycomb lattice of dielectric
 cylinders embedded in a background substance.  
The  honeycomb lattice consists of two inter-penetrating triangular
lattices (called A and B sub-lattices) with the same lattice constant. 
In PhC it is not rare to have the Dirac cone in the
dispersion diagram. Triangular and honeycomb lattices of identical 
circular rods support multiple Dirac cones at the corner of the first 
Brillouin zone. It should be noted that they have the same point group
of six-fold symmetry.  
Doubly degenerate modes at the corner of the first Brillouin zone 
 exhibit the Dirac cone owing to the point group symmetry.\cite{chong2008etq} 
This fact suggests that, the symmetry is crucial and  
the Dirac cone is not limited in the tight-binding model of electrons on
the honeycomb lattice.

Some perturbation breaks the symmetry of the original honeycomb lattice
and causes a crucial influence on the linear dispersion.
In electronic systems the energy difference between A- and B-site atomic orbitals,\cite{semenoff1984cms} periodic
magnetic flux of zero average,\cite{haldane1988mqh} and Rashba spin-orbit interaction\cite{kane2005qsh} are such examples   of the symmetry breaking.  
They break at least either of time-reversal
symmetry (TRS) or space-inversion symmetry (SIS) or parities in
plane. Therefore, the point group of the original honeycomb lattice is
reduced into a smaller group.   
As a result, the two-dimensional irreducible 
representations are prohibited, and the doubly-degenerate modes 
are lifted. The gap between the lifted modes is correlated with  
the magnitude of the symmetry breaking. The effective theory around a
nearly degenerate point is described by the massive Dirac Hamiltonian,
where the mass gap can be controlled via the degree of the symmetry breaking.

In the honeycomb lattice PhCs the TRS is efficiently broken by
applying a magnetic 
field parallel to the cylindrical axis. Nonzero static magnetic field 
 induces 
imaginary off-diagonal elements in the permittivity or permeability 
tensors, through the magneto-optical effect. 
The SIS is broken if the A-site rods are
different from the B-site rods.\cite{onoda-ochiai_short}  
Therefore, we can continuously tune  
the degree of the symmetry breaking in the honeycomb lattice PhCs. 
This tunability is a great advantage of the photonic analog of
graphene model and its extension.\cite{haldane1988mqh} 
From a theoretical point  of view, the tight-binding approximation,
which is commonly used in modeling of graphene, is not widely applicable
for photonic band calculation. Accordingly, the non-bonding orbital in the
nearest neighbor tight-binding approximation, which is responsible for
flatness of the dispersion curve of the zigzag edge state,  is
completely absent in PhCs.   
For example, even in the original honeycomb PhC with both the TRS and the
SIS, the dispersion of the zigzag edge states is not flat because of the
absence of the non-bonding orbital.

Regarding the system with boundary, photonic system is quite distinct
from electronic system. In the latter system the electrons near Fermi
level are prohibited to escape to the outer region via the work
function, {\it i.e.}, a confining potential, and the wave functions of the electrons
are evanescent in the outer region. Therefore, to sustain an edge
state, formation of the band gap in bulk is the minimum requirement. 
On the other hand, in the former system  confining potentials for
photon are absent at the boundary. 
Energy of photon is always positive as in free
space, and no energy barrier exists between the PhC and free
space. 
The simplest way to confine photonic edge states in the PhC is to utilize the light
cone. This restriction of the confinement
makes photonic systems
quit nontrivial in various aspects. 
In particular, the topological relation
between bulk and edge\cite{hatsugai1993cna} 
is of high interest in photonic systems without TRS. 
In quantum Hall system nontrivial topology of bulk states leads to the emergence of chiral edge
states, which are robust against localization effect. 
The edge states play a crucial
role in this system.\cite{halperin1982qhc,wen1991gbe}  Recently, Haldane
and Raghu proposed one-way light waveguide realized in PhCs without
TRS.\cite{haldane2008prd}  Explicit construction of such waveguides is demonstrated by 
several authors.\cite{wang2008rfo,yu2008owe,raghu2008analogs,takeda2008compact}  
This paper also shades light to this topic, by using simpler structure
than those demonstrated so far.

This paper is organized as follows. Section II is devoted to present
bulk properties of the PhC with and/or without TRS and SIS. A numerical
method to deal with edge states is given in Sec. III. Properties of
zigzag and armchair edge states are investigated in detail in Secs. IV
and V, respectively. A one-way light transport along
the edge of a rectangular-shaped PhC is demonstrated in Sec. VI.  
Finally, summary and discussions are given in Sec. VII.

\section{Dirac cone and band gap}
\begin{figure}[h]
\includegraphics*[width=0.45\textwidth]{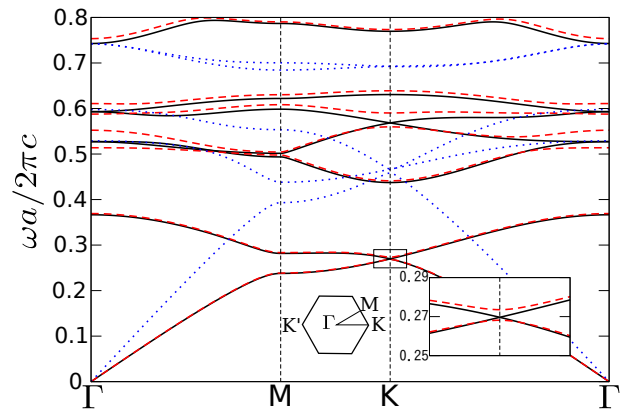}
\caption{\label{band} (Color online) The photonic band structure of the honeycomb 
lattice PhCs of dielectric cylinders embedded in air. 
Solid (dashed) line stands for the TM band structure of 
 the PhC with (without) TRS. The SIS holds in both the cases. 
The dielectric constant and radius of the cylinders 
are taken to be 12 and $0.2a$, respectively, where $a$ is the lattice
 constant.  The magnetic permeability  
of the cylinders 
is taken to be 1 for the PhC with TRS and is given by Eq. (\ref{rodmu})
for the PhC without TRS.  For comparison, the TE band structure of the
 PhC with TRS is shown by doted line. 
}
\end{figure}

Let us consider two-dimensional PhCs composed of the 
honeycomb array of circular cylinders embedded in air. 
The photonic band structure of the PhCs with and without TRS is shown in
Fig. \ref{band} for the transverse magnetic (TM) polarization. 
For comparison, the photonic band structure of the transverse electric
(TE) polarization is also shown for the PhC with TRS. 
The SIS holds in all the cases.   
Here, the dielectric constants $\varepsilon_{A(B)}$ and radius $r_{A(B)}$ of 
the A(B)-cylinders  are taken to be 12 and $0.2a$, respectively. 
The magnetic permeability of the cylinders is taken to be 1 for the PhC 
with TRS, and
has the tensor form given by 
\begin{eqnarray}
\hat{\mu}=
\left(
\begin{array}{ccc}
\mu & i\kappa & 0 \\
-i\kappa & \mu & 0 \\
0 & 0 & \mu
\end{array}
\right),\quad \mu=1, \quad \kappa=0.2, \label{rodmu}
\end{eqnarray}
for the PhC without TRS. 
The first, second, and third rows (columns) stand for $x,y$ and $z$
Cartesian components, respectively. 
The cylindrical axis is taken to be parallel to the $z$ axis. 
The imaginary off-diagonal components of $\hat{\mu}$ are responsible for the
magneto-optical effect and break the TRS. Thus, parameter $\kappa$
represents the degree of the TRS breaking.

As mentioned in Introduction, for the PhC with TRS 
the Dirac cone is found at the K point.  In particular, the first (lowest) and  second
TM bands are in contact with each other at the K point.  They are also in
contact with the K' point because of the spatial symmetry. This property
is quite similar to the tight-binding electron in graphene. As for  
the Dirac point at $\omega a/2\pi c\simeq 0.55$ of the TM polarization, 
the fourth band is in
contact with the fifth band at K (and K'), whereas the former and the
latter are  also in
contact with the third and sixth bands, respectively at the $\Gamma$
point.  
Concerning the TE polarization, the Dirac cones are not clearly visible, 
but are indeed formed between the second and third and between the
fourth and fifth.

On the other hand, in the PhC without TRS, all the degenerate modes at
$\Gamma$ and K are lifted.  The point group of this PhC becomes $C_6$ and the point
group of ${\boldsymbol k}$ at the K point is $C_3$. They are abelian groups, allowing
solely one-dimensional representations. Therefore, the degeneracy is
forbidden.  The energy gap between the lifted modes is proportional to 
$\kappa$ if it is small enough.   
The SIS breaking, $\Delta\varepsilon=\varepsilon_A-\varepsilon_B$, lifts
the double degeneracy at K, but not at $\Gamma$ when the TRS is
preserved.  The energy gap between the lifted modes is proportional to 
$\Delta\varepsilon$.\cite{onoda-ochiai_short}

\begin{figure}[h] 
\includegraphics*[width=0.45\textwidth]{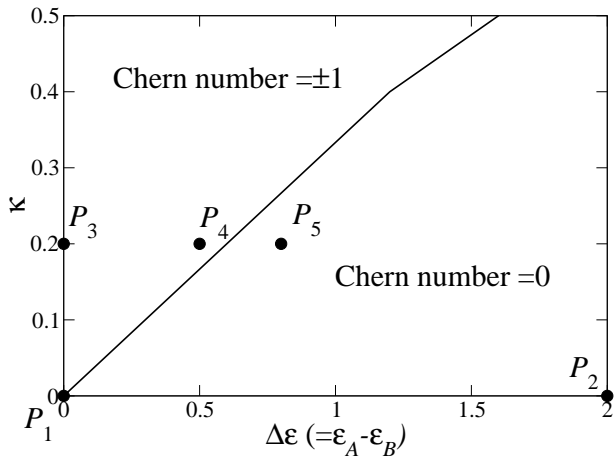}
\caption{\label{phase}  Phase diagram of the honeycomb lattice PhCs for the TM
 polarization. Phase space is spanned by two parameters, $\Delta\epsilon$
 and $\kappa$, which represent the SIS and TRS breaking, respectively.  
The average dielectric constant  
and the radius of the cylinders are kept fixed to $(\varepsilon_A+\varepsilon_B)/2=12$ and $r_A=r_B=0.2a$,
 respectively.  
}
\end{figure}

Let us focus on the gap between the first and second TM bands of the PhC
as a function of the SIS and TRS breaking parameters.  
The phase diagram of the PhC concerning the gap 
is shown in Fig. \ref{phase}. 
At generic values of the parameters the gap opens. However, if we change 
the parameters along a certain curve in the parameter space, the gap
remains to close as shown in Fig. \ref{phase}.  
This property implies
that at finite $\kappa$ the gap closes only at a certain value of
$\Delta\varepsilon$. 
Although the gap opens in both the regions above and below the curve, 
the two regions are topologically different, and are characterized by 
the Chern numbers of the first and second photonic bands. 
The Chern number is a topological integer defined by 
\begin{eqnarray}
& &C_n=\frac{1}{2\pi}\int_{\rm BZ} d^2k (\nabla_{\boldsymbol k} \times {\boldsymbol
 \Lambda}_{n{\boldsymbol k}})_z, \\
& &{\boldsymbol \Lambda}_{n{\boldsymbol k}}=-i\langle
u_{n{\boldsymbol k}}|\nabla_{\boldsymbol k}|u_{n{\boldsymbol k}}\rangle, \\
& &\langle u_{m{\boldsymbol k}}|u_{n{\boldsymbol k}}\rangle=\frac{1}{A}\int_{\rm UC} d^2x
u_{m{\boldsymbol k}}({\boldsymbol x}) \varepsilon({\boldsymbol x})
u_{n{\boldsymbol k}}({\boldsymbol x})=\delta_{m,n} 
\end{eqnarray}
for each non-degenerate band. 
Here, BZ, UC, and $A$ stand for Brillouin zone, unit cell, and the area
of unit cell, respectively. 
The envelop function $u_{n{\boldsymbol k}}({\boldsymbol x})$ of the
$n$-th Bloch state at ${\boldsymbol k}$ is of $E_z$ ({\it i.e.}, the $z$
component of the electric field). 
In the upper region of Fig. \ref{phase},  
$C_1=-1$ and $C_2=1$, whereas in the lower region 
 $C_1=C_2=0$.  
At the gap closing point, the Chern number transfers between  the upper
and lower band under the topological number conservation law.\cite{avron1983haq}  
We will see that the phase diagram correlates with a 
property of edge states in corresponding PhC stripes.  
This correlation is a guiding
principle to design a one-way light transport near PhC edges.\cite{haldane2008prd} 
   
Figure \ref{phase} shows solely the phase diagram in the first quadrant of real
$\Delta\varepsilon$ and $\kappa$. 
The mirror reflection with respect to the $\Delta\varepsilon$ axis gives the phase
diagram in the fourth quadrant, where   $C_1$ and $C_2$ 
are interchanged due to the inversion of $\kappa$.
The phase diagram in the second and third quadrants is obtained by the
mirror reflection with respect to the $\kappa$ axis. The resulting phase
diagram is similar to that obtained in a triangular lattice PhC with anisotropic 
rods.\cite{raghu2008analogs}

\section{characterization of  edge states}
So far, we have concentrated on properties of the PhCs of infinite
extent in plane. If the system has edges, there can emerge edge states
which  are
localized near the edges and  are evanescent 
both inside and outside the PhC.  
In this section  
we introduce a PhC stripe with two parallel boundaries. 
  The boundaries are supposed to have infinite extent, so that the
translational invariance along the boundary still holds. 
The edge states are characterized by Bloch wave vector 
parallel to the boundary.

Optical properties of the PhC stripe are described by the S-matrix. 
It relates the incident plane wave of parallel momentum $k_\|+G'$ to the outgoing
plane wave of  parallel momentum $k_\|+G$, where $G$ and $G'$ are the
reciprocal lattice vectors relevant to the periodicity parallel to the
stripe.\cite{Pendry-LEED-book} 
Both the waves can be evanescent. 
To be precise, the S-matrix is defined by
\begin{eqnarray}
\left(
\begin{array}{c}
(a_+^{\rm out})_G \\
(a_-^{\rm out})_G 
\end{array}
\right)=\sum_{G'}
\left(
\begin{array}{cc}
(S_{++})_{GG'} & (S_{+-})_{GG'} \\
(S_{-+})_{GG'} & (S_{--})_{GG'}
\end{array}
\right)
\left(
\begin{array}{c}
(a_+^{\rm in})_{G'} \\
(a_-^{\rm in})_{G'} 
\end{array}
\right),
\end{eqnarray}
where  $(a_\pm^{\rm in(out)})_{G}$ is the plane-wave-expansion
components of upward (+) and downward (-) incoming (outgoing) waves of 
parallel momentum $k_\|+G$, respectively. 
In our PhCs under consideration the S-matrix  can be calculated via
the photonic layer-Korringa-Kohn-Rostoker
method\cite{Ohtaka:U:A::57:p2550-2568:1998}   as a function of parallel
momentum $k_\|$ and frequency $\omega$.
If the S-matrix is numerically available, the dispersion relation 
of the edge states is obtained according to the following secular equation: 
\begin{eqnarray}
0={\rm det}[S^{-1}].
\end{eqnarray}
Strictly speaking, this equation also includes solutions of bulk states below
the light line. If we search for the solutions inside pseudo gaps ({\it
i.e.} $k_\|$-dependent gaps), 
solely the dispersion relations of the edge states  are obtained. 
In actual calculation, however,  the magnitude
of ${\rm det}[S]$ becomes extremely small with
increasing size of the matrix.  The matrix size is given by the number
of reciprocal lattice vectors taken into account in numerical
calculation. 
In order to obtain numerical accuracy, we have to deal with
larger matrix. 
Therefore, this procedure to determine the edge states 
is generally unstable. Instead, we employ the following scheme. 
Suppose that the S-matrix is divided into two parts $S^u$ and $S^l$ that
correspond to the division of the PhC stripe into the upper and lower
parts. This division is arbitrary, unless the upper or lower part is
not empty.  The following secular equation also determines the dispersion
relation of the edge states: 
\begin{eqnarray}
0={\rm det}[ 1-S_{-+}^l S_{+-}^u].
\end{eqnarray}
This scheme is much stable for larger matrix.

As far as true edge states are concerned, the secular equation has the
zeros in  real axis of frequency  for a given real $k_\|$. 
Here we should also mention leaky edge states ({\it i.e.}, resonances
near the edges),  which are not evanescent outside the PhC
but are evanescent inside the PhC.  Such an edge state is still meaningful,
because the DOS exhibits a peak there. The peak frequency as a function
of parallel momentum $k_\|$ follows a certain curve that is connected to
the dispersion curve of the true edge states. 
To evaluate the leaky edge states, the method developed by Ohtaka {\it et
al}\cite{Ohtaka:I:Y::70:p035109:2004} is employed. In this method, the
DOS at fixed $k_\|$  and $\omega$ is calculated with the truncated
S-matrix of open diffraction channels. The unitarity of the truncated
S-matrix enables us to determine the DOS via eigen-phase-shifts of the
S-matrix.  A peak of the DOS inside the pseudo gap corresponds to a leaky edge state.

\section{zigzag edge}

\begin{figure}[h] 
\includegraphics*[width=0.45\textwidth]{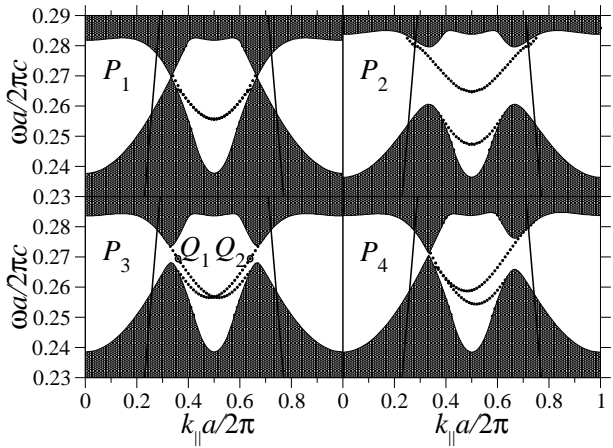}
\caption{\label{pband} The projected band diagrams at point $P_n$ in
the phase diagram (Fig. \ref{phase}) and the dispersion curves of 
the edge states.  The zigzag edge is assumed. 
The shaded regions represent bulk states. The edge states are of the
PhC stripes with 16 layers. Thin solid line stands for the light line. 
 The surface Brillouin zone is taken to
 be $0\le k_\| a/2\pi\le 1$ in order to see the connectivity of the edge-state
 dispersion curves.  
}
\end{figure}

First, let us consider the zigzag edge. 
Figure \ref{pband} shows four sets of the projected band diagram of the honeycomb PhC
and the dispersion relation of the edge states localized near the zigzag
edges.  In Fig. \ref{pband} the shaded regions represent bulk states,
whereas the blank regions correspond to the pseudo gap. Inside the
pseudo gap edge states can emerge. 
In the evaluation of the edge states, we assumed the PhC stripe 
of $N=16$, being $N$ the number of the
layers along the direction perpendicular to the zigzag edges.  
Here, we close up the first and second bands. Higher bands 
are well separated from the lowest two bands. 
Each set refers to either of four points indicated in the phase
diagram of Fig. \ref{phase}.   
In accordance with the Dirac cone in Fig. \ref{band}, 
the projected band structure of point $P_1$ also exhibits 
a point contact at $k_\|a/2\pi=1/3$ and $2/3$. 
The first and second bands are separated for $P_2$ and $P_3$, but are nearly in
contact at  $k_\|a/2\pi=1/3$ for $P_4$. This is because $P_4$ is close to the
phase boundary.     
Except for the lower right panel, 
in which the TRS and the SIS are broken, 
the projected band diagrams and the edge-state dispersion curves are symmetric
with respect to $k_\|a/2\pi=0.5$. This symmetry is preserved  if either the TRS or
the SIS holds.  The time-reversal transformation implies 
\begin{eqnarray}
\omega_n(-k_\|,-k_\perp;\Delta\varepsilon,-\kappa)
=\omega_n(k_\|,k_\perp;\Delta\varepsilon,\kappa), \label{T-transform}
\end{eqnarray}
where $\omega_n(k_\|,k_\perp;\Delta\varepsilon,\kappa)$ is the
eigen-frequency of the $n$-th Bloch state at given parameters
 of $\Delta\varepsilon$ and $\kappa$, and $k_\perp$ is the momentum
 perpendicular to the edge.   
Since $\kappa$ is inverted, Eq. (\ref{T-transform}) is not a symmetry of
the PhC, but is just a transformation law.  
In the case of $\kappa=0$, after the projection concerning $k_\perp$,  
the symmetry with respect to $k_\|=0$ is
obtained.  This symmetry combined with the translational invariance
under $k_\|\to k_\|+G$ results in the symmetry with respect to  $k_\|a/2\pi=0.5$.
Similarly, the space inversion results in   
\begin{eqnarray}
\omega_n(-k_\|,-k_\perp;-\Delta\varepsilon,\kappa)
=\omega_n(k_\|,k_\perp;\Delta\varepsilon,\kappa). \label{P-transform}
\end{eqnarray} 
The symmetry with respect to $k_\|a/2\pi=0$ and $0.5$ is obtained at 
$\Delta\varepsilon=0$.

When edge states are well defined in PhCs with enough number of
layers,  their dispersion relation 
satisfies 
\begin{eqnarray}
& &\omega_{e_1(e_2)}(-k_\|;\Delta\varepsilon,-\kappa)=
\omega_{e_1(e_2)}(k_\|;\Delta\varepsilon,\kappa), \label{T-transform_edge}\\
& &\omega_{e_1}(-k_\|;-\Delta\varepsilon,\kappa)=
\omega_{e_2}(k_\|;\Delta\varepsilon,\kappa),\label{P-transform_edge}
\end{eqnarray}
owing to the time-reversal and space-inversion transformations, respectively. 
Here, $\omega_{e_1}$ and $\omega_{e_2}$ denote the dispersion relation
of opposite edges of the PhC stripe.  
At $\kappa=0$, both $\omega_{e_1}$ and $\omega_{e_2}$ are symmetric
under the inversion of $k_\|$. In contrast, at $\Delta\varepsilon=0$ they are
interchanged. The resulting band diagram is symmetric with respect to
$k_\|a/2\pi=0$ and $0.5$ as in Fig. \ref{pband}.

The upper left panel of Fig. \ref{pband} shows two almost-degenerate
curves that are lifted a bit near the Dirac point. 
This lifting comes from the hybridization between edge states of the
opposite boundary, owing to finite width of the stripe. 
The lifting becomes smaller with increasing $N$, and
eventually two curves merge with each other.  Since $P_1$ corresponds to
$\Delta\varepsilon=\kappa=0$, we obtain $\omega_{e1}=\omega_{e2}$
owing to Eqs. (\ref{T-transform_edge}) and (\ref{P-transform_edge}),  
irrespective of $k_\|$.
As is the same with in graphene, our edge states appear only in the region 
$1/3 \le k_\|a/2\pi \le 2/3$. 
However, the edge-state curves are not flat, in a striking
contrast to the zigzag edge state in the nearest-neighbor tight-binding model of
graphene.

In the upper right panel two edge-state curves are separated in frequency and
each curve terminates in the same bulk band.  On the contrary, in the lower two
panels the dispersion curves of the two edge states intersect one another 
at a particular point
and each curve terminates at different bulk bands. For instance, in the
lower left panel, the curve including $Q_1$ terminates at the upper band 
near $k_\|a/2\pi=1/3$ and at the  lower band near $k_\|a/2\pi=2/3$. 
At other points in the parameter space, we found that the two edge-state
curves are separated if the system is
in the phase of zero Chern number.  Otherwise, if the system is
in the phase of non-zero Chern number, the two curves intersect one another.

\begin{figure}[h]
\begin{center} 
\includegraphics*[width=0.5\textwidth]{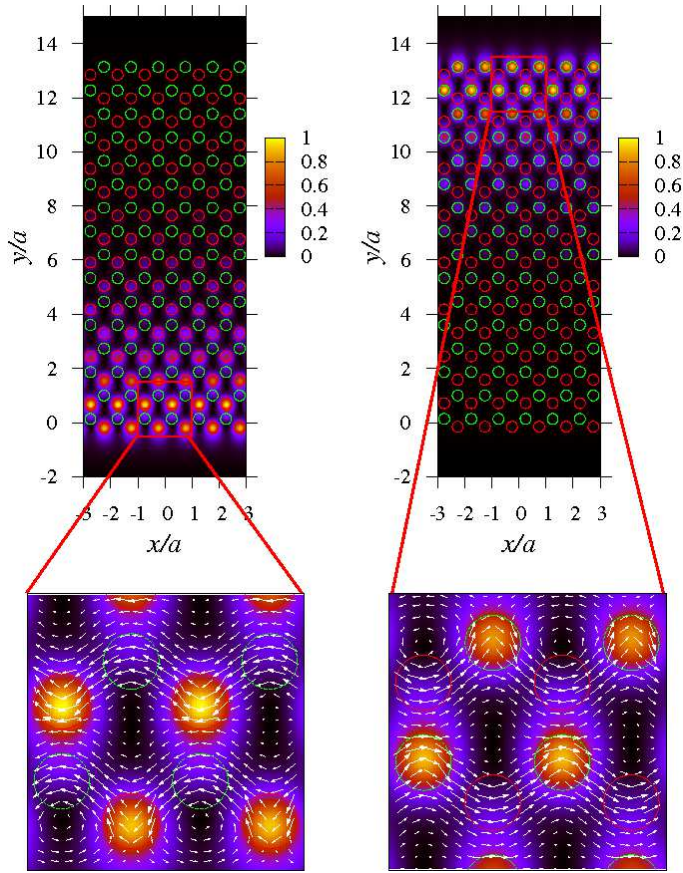}
\caption{\label{edge} (Color online) The electric field intensity $|E_z|^2$  of the
 true edge state at $Q_1$ (left panel) and $Q_2$ (right panel) 
in Fig. \ref{pband}.  
The intensity maxima is normalized as 1. In the enlarged panels 
the Poynting vector flow is also shown. 
}
\end{center}
\end{figure}

The wave function of the edge state at marked points $Q_1$ and $Q_2$ 
is plotted in Fig. \ref{edge}.  
We can easily see that the edge states at $Q_1$ and $Q_2$ are localized
near different edges. This property is consistent with the fact that 
at $\Delta\varepsilon=0$,  
$\omega_{e_1}$ and $\omega_{e_2}$ are interchanged under the inversion of
$k_\|$.
The field configuration at $Q_1$  is
identical to that at $Q_2$ after the space-inversion operation 
($\pi$ rotation). Since the SIS is preserved in this case, they are the
 SIS partners. 
It is also remarkable that the electric field intensity is confined
almost in the rods forming one particular sub-lattice.  This field pattern is
reminiscent of the non-bonding orbital  of the zigzag edge state in
graphene.    
The edge state at $Q_1 (Q_2)$ has the negative (positive) group
velocity.
 Moreover, no other bulk and edge states exist at the frequency. Therefore, 
solely the propagation from left to right is allowed near the upper
edge, while the propagation from left to right is allowed in the lower 
edge. 
  In this way a one-way light transport is realized near a given 
edge.  The one-way transport is robust against quenched disorder with
long correlation length, because the edge states are out of the light line
and the
bulk states at the same frequency is completely absent.\cite{ono1989ipn} 
This is also the case in the lower right panel of Fig. \ref{pband}, 
although the frequency range of the one-way transport is very narrow. 
It should be noted that the non-correlated disorder would cause the
scattering into the states above the light line, where the energy
leakage takes place. Detailed investigation of disorder effects 
is beyond the scope of the present paper.

The results obtained in this section strongly support the bulk-edge 
correspondence, which was originally proven in the context of quantum
Hall systems\cite{hatsugai1993cna} and was discussed in the context of photonic systems recently.\cite{haldane2008prd} 
Namely, the number of one-way edge states in a given two-dimensional 
omni-directional gap
({\it i.e.} $k_\|$-independent gap)  
is equal to the sum of the Chern numbers of the bulk bands
below the gap.  In our case 
the Chern number of the lower (upper) band is equal to -1
(1). A negative sign of the sum corresponds to the inverted direction of
the edge propagation.  
Accordingly, there is only one (one-way) state per edge in the
gap between the first and second bands. 
Moreover, no edge state is found between the second and third bands. 
This behavior is consistent with the Chern numbers of the first and
second bands, according to the bulk-edge correspondence.

Finally, let us briefly comment on the edge states in $P_1$ and $P_2$. 
In $P_1$ the two edge states are completely degenerate at $N=\infty$. 
For the system with narrow width, there appear 
the bonding and anti-bonding orbitals,  
each of which has an equal weight of the field intensity 
in both the zigzag edges.  
As for the edge states in $P_2$,
the upper (lower) edge states are localized near the upper (lower) zigzag
edge.

\section{armchair edge}
\begin{figure}[h] 
\includegraphics*[width=0.45\textwidth]{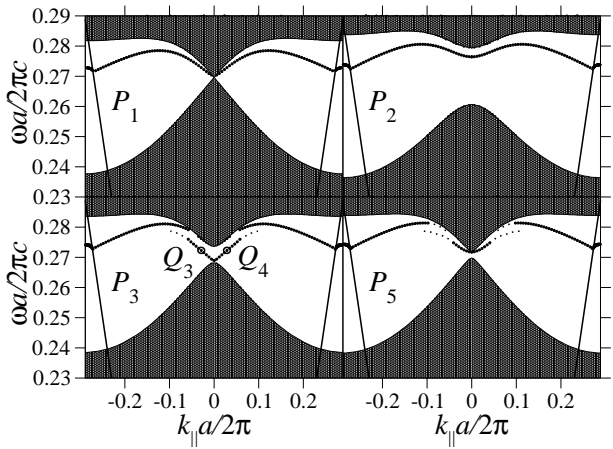}
\caption{\label{pband_armchair} The projected band diagrams at point $P_n$ in
the phase diagram (Fig. \ref{phase}) and the dispersion curves of 
the edge states.  The armchair edge is assumed. 
The shaded regions correspond to bulk states. The edge states are of the
PhC stripes with 64 layers. Thin solid line stands for the light line. 
}
\end{figure}

Next, let us consider the armchair edge. 
The projected band diagram and the dispersion
curves of the edge states are shown in Fig. \ref{pband_armchair}. 
We assumed the PhC stripe with $N=64$. 
It should be noted that they are symmetric with respect to $k_\|=0$
regardless of SIS and TRS.  
This property is understood by the combination of a 
parity transformation  and Eq. (\ref{T-transform}). 
Under the parity transformation with respect to the mirror plane parallel
to the armchair edges, 
\begin{eqnarray}
\omega_n(k_\|,-k_\perp;\Delta\varepsilon,-\kappa)=\omega_n
(k_\|,k_\perp;\Delta\varepsilon,\kappa). \label{Parity}
\end{eqnarray} 
By combining Eqs. (\ref{T-transform}) and (\ref{Parity}), we obtain the symmetric projected band
diagram with respect to $k_\|=0$. 
Concerning the edge states, the parity transformation results in  
\begin{eqnarray}
& &\omega_{e_1}(k_\|;\Delta\varepsilon,-\kappa)=\omega_{e_2}
(k_\|;\Delta\varepsilon,\kappa). \label{Parity_edge}
\end{eqnarray} 
Therefore, by combining  Eqs. (\ref{T-transform_edge}) and 
(\ref{Parity_edge}), we can derive that $\omega_{e_1}$ and 
$\omega_{e_2}$ are interchanged by the inversion of $k_\|$, 
regardless of SIS and TRS:
\begin{eqnarray}
\omega_{e_1}(-k_\|;\Delta\varepsilon,\kappa)=\omega_{e_2}
(k_\|;\Delta\varepsilon,\kappa). \label{interchange}
\end{eqnarray}
Equation (\ref{interchange}) results in the degeneracy between $\omega_{e_1}$ and $\omega_{e_2}$ 
at the boundary of the surface Brillouin zone.   Moreover, it is obvious from
Eq. (\ref{Parity_edge}) that the two edge states are completely degenerate
at $\kappa=0$ in the entire surface Brillouin zone.

In the armchair projection the K and K' points in the first Brillouin
zone are mapped on the same point
$k_\|=0$ in the surface Brillouin zone, being above  the light line. 
Therefore, possible edge states relevant to the Dirac cone are leaky, 
unless the region outside the PhC is screened. 
Accordingly, the DOS of an armchair edge state at fixed $k_\|$ 
shows up as a Lorentzian peak,  in a striking contrast to that of a
zigzag edge state being a delta-function peak.    
The dispersion relation of the leaky edge states depends strongly on the
number of layers $N$.  However, if $N$ is large enough, the $N$-dependence
disappears.  We found that at large enough $N$, the leaky edge states 
correlate with the Chern number  fairly well.

In the case as $P_3$ where the Chern numbers of the upper and lower bands are nonzero, we found
a segment of the dispersion curve of the leaky edge state 
whose bottom is at the lower band edge, as shown in the lower left
panel of Fig. \ref{pband_armchair}.  There also appear another segment of
the dispersion curve which crosses the light line. 
Across the phase boundary, the upper band touches to and
separates from the lower band. After the separation as the case $P_5$, 
the bottom of the 
former segment moves from the lower band edge to the upper band edge
as shown in the lower right panel of Fig. \ref{pband_armchair}.  
By increasing $\Delta\varepsilon$, this segment hides among the upper
bulk band (not shown).  
We should note that the dispersion curve
of the leaky edge states is obtained by
tracing the peak frequencies of the DOS as a function of $k_\|$. 
If a peak becomes a shoulder, we stopped tracing the
curve and indicated  shoulder frequencies by dotted curve. 
This is the case for $P_3$ and $P_5$. 
For $P_3$, the DOS changes its shape from peak
to shoulder at $k_\|a/2\pi \simeq \pm 0.058$. This is why 
the segment including $Q_3$ and $Q_4$ seems to terminate around there.  
However, we can distinguish this shoulder in the region $0.058 <
|k_\||a/2\pi < 0.1$, accompanying an additional peak above it. 
The peak bringing the shoulder with it becomes an asymmetric peak 
for $|k_\||a/2\pi > 0.1$ and crosses the light line. 
In the DOS spectrum of $P_5$, we can find two shoulders just below the
peaks of bulk states in the region $0.04 < |k_\||a/2\pi < 0.1$. 
Again, they merge each other and become an asymmetric peak for $|k_\||a/2\pi > 0.1$.
Such an asymmetric peak consists of two peaks with different heights and
widths, which 
come from the lifting of the degenerate edge
states in the limit of $\kappa=0$. 
Actually, for $P_1$ and $P_2$ in which the edge states are
doubly-degenerate, we can see a nearly-symmetric single peak for the
leaky edge states in each case.

As in the case of zigzag edge, the leaky edge states in the
two-dimensional omni-directional gap 
exhibit a one-way light transport if the relevant Chern number is nonzero.
Here, we consider the structure with two horizontal armchair edges.  
The incident wave with positive $k_\|$ coming from the bottom  
cannot excite the leaky edge state just above the lower band
edge, {\it e.g.},  state $Q_4$ in Fig. \ref{pband_armchair}.
 However,  the incident
wave with negative $k_\|$ coming from the bottom  can excite the
leaky edge state, {\it e.g.}, at $Q_3$.    
In the latter case, the leaky edge state has negative
group velocity, traveling from right to left.  
This relation becomes inverted for the plane wave coming
from the top. The incident plane wave with positive (negative) $k_\|$
can (cannot) excite the leaky edge state localized near the upper
armchair edge. This edge state has positive group velocity, traveling
from left to right.  In this way, one-way light transport is realized as
in the zigzag edge case. 
Under quenched disorder the one-way transport is protected against the
mixing with bulk states, because no bulk state exists in  the
omni-directional gap. 
However, in contrast to the zigzag edge case, 
even the disorder with long correlation length 
could enhance the energy leakage to the outer region.

\begin{figure}[h]
\begin{center}
\includegraphics*[width=0.5\textwidth]{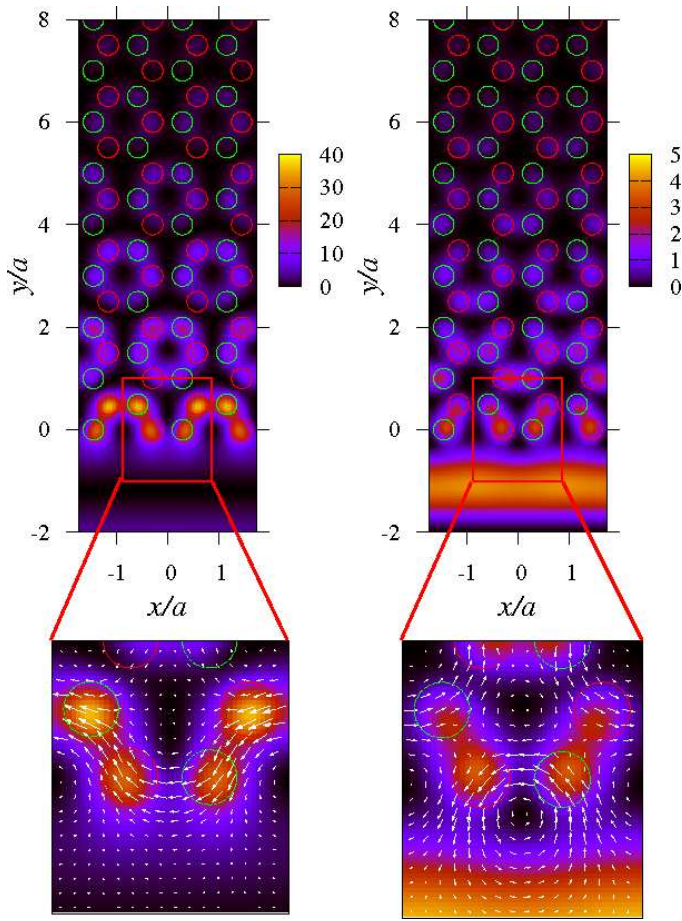}
\caption{\label{gkedgeconf}  (Color online) The electric field intensity $|E_z|^2$
 induced by the incident plane wave having $(k_\|,\omega)$ at $Q_3$
 (left panel) and
 $Q_4$ (right panel) in Fig. \ref{pband_armchair}. 
The incident plane wave of unit intensity comes from the bottom 
of the structure. In the enlarged panels the Poynting vector flow is
 also shown. }
\end{center}
\end{figure}

Figure \ref{gkedgeconf} shows the electric field intensity $|E_z|^2$ 
induced by the incident plane wave whose $\omega$ and $k_\|$ are at the
marked points ($Q_3$ and $Q_4$) in Fig. \ref{pband_armchair}. 
The intensity of the incident plane wave is taken to be 1 and the field
configuration above $y/a=8$ is omitted. 
Although, the dispersion curve is symmetric with
respect to  $k_\|=0$, the field configuration is quite asymmetric. Of
particular importance is the near-field pattern around the lower edge. 
In the left panel the strongest field intensity of order 40 is found in the
boundary armchair layer, whereas in the right panel it is found 
outside  the PhC with much smaller intensity. In both the cases, the
transmittances are the same and nearly equal to zero. Accordingly, no 
field enhancement is observed near the upper edge (not shown). 
The remarkable contrast of the field profiles indicates that the leaky
edge state with horizontal energy flow is excited in the left panel, but is not in the right panel. 
If the plane wave is incident from the top, the field pattern exhibits 
an opposite behavior. 
That is, the plane wave with $\omega$ and $k_\|$ at $Q_4$ from the top  excites the leaky edge
state localized near the upper edge, but at $Q_3$ it cannot excite the
leaky edge state.

The property of each edge state is also understood as follows.  
When we scan $k_\|$ from negative to positive along the dispersion curve of
the leaky edge state, 
the localized center of the edge state transfers from one edge to the other. 
The critical point is at the bottom of the dispersion curve, where 
the edge state merges to the bulk state of the lower band. It is 
extended inside the PhC, making a bridge from one edge to the other. 
The entire picture is consistent with the interchange of $\omega_{e_1}$ and 
$\omega_{e2}$ under the inversion of $k_\|$.

Finally, let us comment on the field configuration of other edge states. 
For $P_1$ and $P_2$, the edge states are degenerate between the upper
and lower edges. Accordingly, the incident plane wave coming from
the bottom (top) of the structure excites the leaky edge states localized
around the bottom (top) edge. It is regardless of the sign of $k_\|$. 
For $P_3$ and $P_5$, the edge-state curve that crosses the light line 
corresponds to an asymmetric peak in the DOS, which is actually the sum
of two peaks.  
It is difficult to separate the two peaks, because they are overlapped 
in frequency. Thus, the edge states can be excited  by the
incident wave coming from both top and bottom of the PhC. 
Concerning the quadratic edge state around $k_\|=0$ of $P_5$, a similar
contrast in the field configuration between positive and negative $k_\|$
is obtained as in $Q_3$ and $Q_4$.  However, under quenched disorder  
this edge state readily mixes with bulk states that exist at the same
frequency.

\section{Demonstration of one-way light transport}

The direction of the one-way transport in the zigzag edge 
is consistent with that in the armchair edge. 
Let us consider a rectangular-shaped PhC whose four
edges are  zigzag, armchair, zigzag, and armchair in a clockwise order. 
The one-way transport found in Figs. \ref{pband} and
\ref{pband_armchair} must be 
clockwise in this geometry. 

To verify it certainly happens, we performed
a numerical simulation of the light transport in the rectangular-shaped 
PhC.  The multiple-scattering method is employed along with a Gaussian
beam incidence.\cite{Bravo-Abad:O:S::67:p115116:2003} 
We assume $N=32$ for the zigzag edges and $N=64$ for the armchair edges.  
The incident Gaussian beam is focused at the midpoint of the
front armchair edge. 
The electric field intensity $|E_z|^2$ at the focused point is normalized
as 1 and the beam waist is $10a$.  
The frequency and
the incident angle of the beam are taken to be $\omega a/2\pi c=0.273$
and $\theta_0=7.263^\circ$, which corresponds to the leaky edge state very close to
 the $Q_3$ point.  The beam waist
size  is chosen to avoid possible diffraction at the corner
of the PhC and not to excite the states near the  $Q_4$ point at the same time. 

\begin{figure}[h]
\begin{center}
\includegraphics*[width=0.45\textwidth]{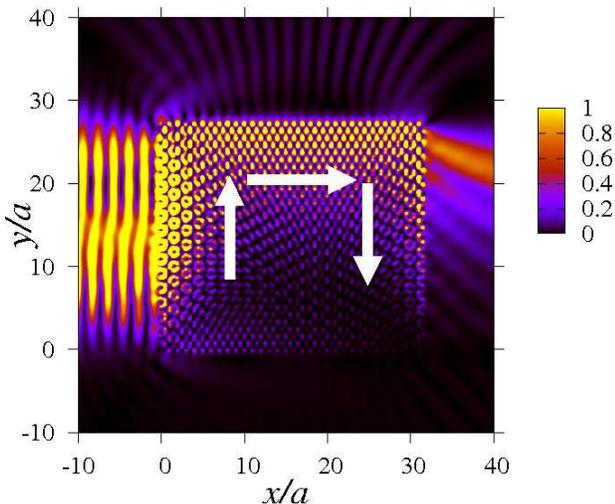}
\caption{\label{onewayconf}  (Color online) The electric field intensity $|E_z|^2$
 induced by the time-harmonic Gaussian beam  coming from the left of
 the rectangular-shaped PhC. The four edges are either zigzag (top and
 bottom) or armchair (left and right). The beam is focused at the mid
 point of the left armchair edge with the unit electric field intensity 
  and beam waist of $10a$.  }
\end{center}
\end{figure}

Resulting electric field intensity $|E_z|^2$ is plotted in
Fig. \ref{onewayconf}. 
The incident beam is almost reflected at the left (armchair) edge, forming the 
interference pattern in the left side of the PhC. However, as in the
left panel of Fig. \ref{gkedgeconf}, the leaky edge state is certainly 
excited there. This edge state propagates upward, and is diffracted at 
the upper left corner. 
A certain portion of the energy turns into the zigzag edge state localized
near the upper edge. This edge state propagates from left to right. 
 The energy leakage at the upper edge is very small compared to that in
the left and right edges. 
This zigzag edge state is more or less diffracted at the upper
right corner. However, the down-going armchair edge state  is
certainly excited  in the right edge. Obviously, the field intensity of
the right edge reduces with reducing $y$ coordinate. 
This behavior is consistent with the energy leakage of the armchair edge
state. Finally, the field intensity almost vanished at the lower right
corner.  In this way, the clockwise  one-way light transport is realized 
in the rectangular-shaped PhC. 

We also confirmed that the incident beam
with the same parameters but inverted incident angle ($-\theta_0$) does
not excite the counterclockwise one-way transport along the edges. 
The incident beam is just reflected without exciting the relevant leaky
edge state  in accordance with the right panel of Fig. \ref{gkedgeconf}.

\section{Summary and discussions}   
In summary, we have presented a numerical analysis on the bulk and 
edge states in honeycomb lattice PhCs as a photonic analog of
graphene model and its extension.  
In the TM polarization the Dirac cone emerges 
between the first and second bands. 
The mass gap in the Dirac cone is controllable by the
parameters of the SIS or TRS breaking. On a certain curve in the
parameter space, the band touching takes place. This curve divides 
the parameter space into two topologically-distinct regions. One is
characterized by zero Chern number of the upper and lower bands, and the
other is characterized by Chern number of $\pm 1$.  
Of particular
importance is the correlation between the Chern number in bulk and
light transport near edge. Non-zero Chern number in
bulk photonic bands 
results in one-way light transport near the edge. 
It is quite similar
to the bulk-edge correspondence found in quantum Hall systems. 

In this paper we focus on the TM polarization in rod-in-air type PhCs. 
This is mainly because the band touching takes place between the lowest two
bands and they are well separated from  higher bands by the wide
band gap, provided that the refractive index of the rods are high
enough.   
 In rod-in-air type PhCs the TE polarization results in the band
touching between the second and third  bands. However, the Dirac cone is
not clearly visible, although it is certainly formed.   As for
hole-in-dielectric type PhCs, an opposite tendency is found. Namely, 
the band touching between the lowest two bands takes place only in the
TE polarization. In this case the distance between the boundary column of
air holes and the PhC edge affect edge states. 
Therefore, we must take account of this parameter to determine
the dispersion curves of the edge states. 

Concerning the TRS breaking, we have introduced imaginary
off-diagonal components in the permeability tensor. 
This is the most efficient way to break the TRS for the TM
polarization. Such a permeability tensor is normally 
not available in visible frequency range.\cite{Landau-EDCM-book} 
However, in GHz range it is possible to obtain $\kappa$ 
of order 10. Such a large $\kappa$ is necessary to obtain a robust one-way
transport against thermal fluctuations, etc.  In the numerical setup we assume 
an intermediate frequency range with smaller $\kappa$.
 On the other hand, in the TE polarization, 
the TRS can be efficiently broken by  imaginary
off-diagonal components in the permittivity tensor. In this case 
the PhC without the TRS can operate in visible frequency range.
However, strong magnetic field is necessary in order to induce 
large imaginary off-diagonal components of the permittivity tensor.  
Thus, it is strongly desired to explorer low-loss optical media with large 
magneto-optical effect, in order to have robust one-way transport. 
         
Recently, another photonic analog of graphene, namely, honeycomb array
of metallic nano-particles,  is proposed and analyzed
theoretically.\cite{han:123904} Particle plasmon resonances in the nano-particles act as
 if localized orbitals in Carbon atom. The tight-binding picture 
 is thus reasonably adapted to this system, and nearly flat
bands are found in the zigzag edge.  Vectorial nature of photon plays a
crucial role there, giving rise to a remarkable feature  in the
dispersion curves of the edge states in the quasi-static approximation.  
In contrast, vectorial nature of
photon is minimally introduced in our model, but a full analysis including
possible  retardation effects and symmetry-breaking effects has been 
made.  Effects of the TE-TM mixing in off-axis propagation are an
important issue in our system. 
In particular, it is interesting to study to what extent the bulk-edge
correspondence is modified. 
We hope this paper stimulates further investigation based on the analogy between
electronic and photonic systems on honeycomb lattices.

\begin{acknowledgments}
The work of T. O. was partially supported by Grant-in-Aid (No. 20560042)
 for Scientific Research from the Ministry of Education, Culture,
 Sports, Science and Technology.   
\end{acknowledgments}

\begin{thebibliography}{27}
\expandafter\ifx\csname natexlab\endcsname\relax\def\natexlab#1{#1}\fi
\expandafter\ifx\csname bibnamefont\endcsname\relax
  \def\bibnamefont#1{#1}\fi
\expandafter\ifx\csname bibfnamefont\endcsname\relax
  \def\bibfnamefont#1{#1}\fi
\expandafter\ifx\csname citenamefont\endcsname\relax
  \def\citenamefont#1{#1}\fi
\expandafter\ifx\csname url\endcsname\relax
  \def\url#1{\texttt{#1}}\fi
\expandafter\ifx\csname urlprefix\endcsname\relax\def\urlprefix{URL }\fi
\providecommand{\bibinfo}[2]{#2}
\providecommand{\eprint}[2][]{\url{#2}}

\bibitem[{\citenamefont{Novoselov et~al.}(2005)\citenamefont{Novoselov, Geim,
  Morozov, Jiang, Katsnelson, Grigorieva, Dubonos, and
  Firsov}}]{novoselov2005tdg}
\bibinfo{author}{\bibfnamefont{K.~S.} \bibnamefont{Novoselov}},
  \bibinfo{author}{\bibfnamefont{A.~K.} \bibnamefont{Geim}},
  \bibinfo{author}{\bibfnamefont{S.~V.} \bibnamefont{Morozov}},
  \bibinfo{author}{\bibfnamefont{D.}~\bibnamefont{Jiang}},
  \bibinfo{author}{\bibfnamefont{M.~I.} \bibnamefont{Katsnelson}},
  \bibinfo{author}{\bibfnamefont{I.~V.} \bibnamefont{Grigorieva}},
  \bibinfo{author}{\bibfnamefont{S.~V.} \bibnamefont{Dubonos}},
  \bibnamefont{and} \bibinfo{author}{\bibfnamefont{A.~A.}
  \bibnamefont{Firsov}}, \bibinfo{journal}{Nature}
  \textbf{\bibinfo{volume}{438}}, \bibinfo{pages}{197} (\bibinfo{year}{2005}).

\bibitem[{\citenamefont{Geim and Novoselov}(2007)}]{geim2007rg}
\bibinfo{author}{\bibfnamefont{A.~K.} \bibnamefont{Geim}} \bibnamefont{and}
  \bibinfo{author}{\bibfnamefont{K.~S.} \bibnamefont{Novoselov}},
  \bibinfo{journal}{Nature Materials} \textbf{\bibinfo{volume}{6}},
  \bibinfo{pages}{183} (\bibinfo{year}{2007}).

\bibitem[{\citenamefont{Klein}(1929)}]{klein}
\bibinfo{author}{\bibfnamefont{O.}~\bibnamefont{Klein}}, \bibinfo{journal}{Z.
  Physik A} \textbf{\bibinfo{volume}{53}}, \bibinfo{pages}{157}
  (\bibinfo{year}{1929}).

\bibitem[{\citenamefont{Schrodinger}(1930)}]{Schrodinger}
\bibinfo{author}{\bibfnamefont{E.}~\bibnamefont{Schrodinger}},
  \bibinfo{journal}{Sitzungsb. Preuss. Akad. Wiss. Phys.-Math. Kl.}
  \textbf{\bibinfo{volume}{24}}, \bibinfo{pages}{418} (\bibinfo{year}{1930}).

\bibitem[{\citenamefont{Nakada et~al.}(1996)\citenamefont{Nakada, Fujita,
  Dresselhaus, and Dresselhaus}}]{nakada1991esg}
\bibinfo{author}{\bibfnamefont{K.}~\bibnamefont{Nakada}},
  \bibinfo{author}{\bibfnamefont{M.}~\bibnamefont{Fujita}},
  \bibinfo{author}{\bibfnamefont{G.}~\bibnamefont{Dresselhaus}},
  \bibnamefont{and} \bibinfo{author}{\bibfnamefont{M.~S.}
  \bibnamefont{Dresselhaus}}, \bibinfo{journal}{Phys. Rev. B}
  \textbf{\bibinfo{volume}{54}}, \bibinfo{pages}{17954} (\bibinfo{year}{1996}).

\bibitem[{\citenamefont{Cassagne et~al.}(1995)\citenamefont{Cassagne, Jouanin,
  and Bertho}}]{cassagne1995pbg}
\bibinfo{author}{\bibfnamefont{D.}~\bibnamefont{Cassagne}},
  \bibinfo{author}{\bibfnamefont{C.}~\bibnamefont{Jouanin}}, \bibnamefont{and}
  \bibinfo{author}{\bibfnamefont{D.}~\bibnamefont{Bertho}},
  \bibinfo{journal}{Phys. Rev. B}, \textbf{\bibinfo{volume}{52}},
  \bibinfo{pages}{R2217} 
  (\bibinfo{year}{1995}).

\bibitem[{\citenamefont{Chong et~al.}(2008)\citenamefont{Chong, Wen, and
  Solja{\v{c}}i{\'c}}}]{chong2008etq}
\bibinfo{author}{\bibfnamefont{Y.~D.} \bibnamefont{Chong}},
  \bibinfo{author}{\bibfnamefont{X.~G.} \bibnamefont{Wen}}, \bibnamefont{and}
  \bibinfo{author}{\bibfnamefont{M.}~\bibnamefont{Solja{\v{c}}i{\'c}}},
  \bibinfo{journal}{Phys. Rev. B} \textbf{\bibinfo{volume}{77}},
  \bibinfo{pages}{235125} (\bibinfo{year}{2008}).

\bibitem[{\citenamefont{Semenoff}(1984)}]{semenoff1984cms}
\bibinfo{author}{\bibfnamefont{G.~W.} \bibnamefont{Semenoff}},
  \bibinfo{journal}{Phys. Rev. Lett.} \textbf{\bibinfo{volume}{53}},
  \bibinfo{pages}{2449} (\bibinfo{year}{1984}).

\bibitem[{\citenamefont{Haldane}(1988)}]{haldane1988mqh}
\bibinfo{author}{\bibfnamefont{F.~D.~M.} \bibnamefont{Haldane}},
  \bibinfo{journal}{Phys. Rev. Lett.} \textbf{\bibinfo{volume}{61}},
  \bibinfo{pages}{2015} (\bibinfo{year}{1988}).

\bibitem[{\citenamefont{Kane and Mele}(2005)}]{kane2005qsh}
\bibinfo{author}{\bibfnamefont{C.~L.} \bibnamefont{Kane}} \bibnamefont{and}
  \bibinfo{author}{\bibfnamefont{E.~J.} \bibnamefont{Mele}},
  \bibinfo{journal}{Phys. Rev. Lett.} \textbf{\bibinfo{volume}{95}},
  \bibinfo{pages}{226801} (\bibinfo{year}{2005}).

\bibitem[{\citenamefont{Onoda and Ochiai}()}]{onoda-ochiai_short}
\bibinfo{author}{\bibfnamefont{M.}~\bibnamefont{Onoda}} \bibnamefont{and}
  \bibinfo{author}{\bibfnamefont{T.}~\bibnamefont{Ochiai}},
  \bibinfo{note}{arXiv:0810.1101}.

\bibitem[{\citenamefont{Hatsugai}(1993)}]{hatsugai1993cna}
\bibinfo{author}{\bibfnamefont{Y.}~\bibnamefont{Hatsugai}},
  \bibinfo{journal}{Phys. Rev. Lett.} \textbf{\bibinfo{volume}{71}},
  \bibinfo{pages}{3697} (\bibinfo{year}{1993}).

\bibitem[{\citenamefont{Halperin}(1982)}]{halperin1982qhc}
\bibinfo{author}{\bibfnamefont{B.~I.} \bibnamefont{Halperin}},
  \bibinfo{journal}{Phys. Rev. B} \textbf{\bibinfo{volume}{25}},
  \bibinfo{pages}{2185} (\bibinfo{year}{1982}).

\bibitem[{\citenamefont{Wen}(1991)}]{wen1991gbe}
\bibinfo{author}{\bibfnamefont{X.~G.} \bibnamefont{Wen}},
  \bibinfo{journal}{Phys. Rev. B} \textbf{\bibinfo{volume}{43}},
  \bibinfo{pages}{11025} (\bibinfo{year}{1991}).

\bibitem[{\citenamefont{Haldane and Raghu}(2008)}]{haldane2008prd}
\bibinfo{author}{\bibfnamefont{F.~D.~M.} \bibnamefont{Haldane}}
  \bibnamefont{and} \bibinfo{author}{\bibfnamefont{S.}~\bibnamefont{Raghu}},
  \bibinfo{journal}{Phys. Rev. Lett.} \textbf{\bibinfo{volume}{100}},
  \bibinfo{pages}{013904} (\bibinfo{year}{2008}).

\bibitem[{\citenamefont{Wang et~al.}(2008)\citenamefont{Wang, Chong,
  Joannopoulos, and Solja{\v{c}}i{\'c}}}]{wang2008rfo}
\bibinfo{author}{\bibfnamefont{Z.}~\bibnamefont{Wang}},
  \bibinfo{author}{\bibfnamefont{Y.~D.} \bibnamefont{Chong}},
  \bibinfo{author}{\bibfnamefont{J.~D.} \bibnamefont{Joannopoulos}},
  \bibnamefont{and}
  \bibinfo{author}{\bibfnamefont{M.}~\bibnamefont{Solja{\v{c}}i{\'c}}},
  \bibinfo{journal}{Phys. Rev. Lett.} \textbf{\bibinfo{volume}{100}},
  \bibinfo{pages}{013905} (\bibinfo{year}{2008}).

\bibitem[{\citenamefont{Yu et~al.}(2008)\citenamefont{Yu, Veronis, Wang, and
  Fan}}]{yu2008owe}
\bibinfo{author}{\bibfnamefont{Z.}~\bibnamefont{Yu}},
  \bibinfo{author}{\bibfnamefont{G.}~\bibnamefont{Veronis}},
  \bibinfo{author}{\bibfnamefont{Z.}~\bibnamefont{Wang}}, \bibnamefont{and}
  \bibinfo{author}{\bibfnamefont{S.}~\bibnamefont{Fan}},
  \bibinfo{journal}{Phys. Rev. Lett.} \textbf{\bibinfo{volume}{100}},
  \bibinfo{pages}{023902} (\bibinfo{year}{2008}).

\bibitem[{\citenamefont{Raghu and Haldane}(2008)}]{raghu2008analogs}
\bibinfo{author}{\bibfnamefont{S.}~\bibnamefont{Raghu}} \bibnamefont{and}
  \bibinfo{author}{\bibfnamefont{F.~D.~M.} \bibnamefont{Haldane}},
  \bibinfo{journal}{Phys. Rev. A} \textbf{\bibinfo{volume}{78}},
  \bibinfo{pages}{033834} (\bibinfo{year}{2008}).

\bibitem[{\citenamefont{Takeda and John}(2008)}]{takeda2008compact}
\bibinfo{author}{\bibfnamefont{H.}~\bibnamefont{Takeda}} \bibnamefont{and}
  \bibinfo{author}{\bibfnamefont{S.}~\bibnamefont{John}},
  \bibinfo{journal}{Phys. Rev. A} \textbf{\bibinfo{volume}{78}},
  \bibinfo{pages}{023804} (\bibinfo{year}{2008}).

\bibitem[{\citenamefont{Avron et~al.}(1983)\citenamefont{Avron, Seiler, and
  Simon}}]{avron1983haq}
\bibinfo{author}{\bibfnamefont{J.~E.} \bibnamefont{Avron}},
  \bibinfo{author}{\bibfnamefont{R.}~\bibnamefont{Seiler}}, \bibnamefont{and}
  \bibinfo{author}{\bibfnamefont{B.}~\bibnamefont{Simon}},
  \bibinfo{journal}{Phys. Rev. Lett.} \textbf{\bibinfo{volume}{51}},
  \bibinfo{pages}{51} (\bibinfo{year}{1983}).

\bibitem[{\citenamefont{Pendry}(1974)}]{Pendry-LEED-book}
\bibinfo{author}{\bibfnamefont{J.~B.} \bibnamefont{Pendry}},
  \emph{\bibinfo{title}{Low Energy Electron Diffraction}}
  (\bibinfo{publisher}{Academic}, \bibinfo{address}{London},
  \bibinfo{year}{1974}).

\bibitem[{\citenamefont{Ohtaka et~al.}(1998)\citenamefont{Ohtaka, Ueta, and
  Amemiya}}]{Ohtaka:U:A::57:p2550-2568:1998}
\bibinfo{author}{\bibfnamefont{K.}~\bibnamefont{Ohtaka}},
  \bibinfo{author}{\bibfnamefont{T.}~\bibnamefont{Ueta}}, \bibnamefont{and}
  \bibinfo{author}{\bibfnamefont{K.}~\bibnamefont{Amemiya}},
  \bibinfo{journal}{Phys. Rev. B} \textbf{\bibinfo{volume}{57}},
  \bibinfo{pages}{2550} (\bibinfo{year}{1998}).

\bibitem[{\citenamefont{Ohtaka et~al.}(2004)\citenamefont{Ohtaka, Inoue, and
  Yamaguti}}]{Ohtaka:I:Y::70:p035109:2004}
\bibinfo{author}{\bibfnamefont{K.}~\bibnamefont{Ohtaka}},
  \bibinfo{author}{\bibfnamefont{J.~I.}~\bibnamefont{Inoue}}, \bibnamefont{and}
  \bibinfo{author}{\bibfnamefont{S.}~\bibnamefont{Yamaguti}},
  \bibinfo{journal}{Phys. Rev. B} \textbf{\bibinfo{volume}{70}},
  \bibinfo{pages}{035109} (\bibinfo{year}{2004}).

\bibitem[{\citenamefont{Ono et~al.}(1989)\citenamefont{Ono, Ohtsuki, and
  Kramer}}]{ono1989ipn}
\bibinfo{author}{\bibfnamefont{Y.}~\bibnamefont{Ono}},
  \bibinfo{author}{\bibfnamefont{T.}~\bibnamefont{Ohtsuki}}, \bibnamefont{and}
  \bibinfo{author}{\bibfnamefont{B.}~\bibnamefont{Kramer}},
  \bibinfo{journal}{J. Phys. Soc. Jpn.} \textbf{\bibinfo{volume}{58}},
  \bibinfo{pages}{1705} (\bibinfo{year}{1989}).

\bibitem[{\citenamefont{Bravo-Abad et~al.}(2003)\citenamefont{Bravo-Abad,
  Ochiai, and S\'{a}nchez-Dehesa}}]{Bravo-Abad:O:S::67:p115116:2003}
\bibinfo{author}{\bibfnamefont{J.}~\bibnamefont{Bravo-Abad}},
  \bibinfo{author}{\bibfnamefont{T.}~\bibnamefont{Ochiai}}, \bibnamefont{and}
  \bibinfo{author}{\bibfnamefont{J.}~\bibnamefont{S\'{a}nchez-Dehesa}},
  \bibinfo{journal}{Phys. Rev. B} \textbf{\bibinfo{volume}{67}},
  \bibinfo{pages}{115116} (\bibinfo{year}{2003}).

\bibitem[{\citenamefont{Landau et~al.}(1985)\citenamefont{Landau, Lifshitz, and
  Pitaevskii}}]{Landau-EDCM-book}
\bibinfo{author}{\bibfnamefont{L.~D.} \bibnamefont{Landau}},
  \bibinfo{author}{\bibfnamefont{E.~M.} \bibnamefont{Lifshitz}},
  \bibnamefont{and} \bibinfo{author}{\bibfnamefont{L.~P.}
  \bibnamefont{Pitaevskii}}, \emph{\bibinfo{title}{Electrodynamics of
  Continuous Media}} (\bibinfo{publisher}{Butterworth-Heinemann},
  \bibinfo{address}{Oxford}, \bibinfo{year}{1985}).

\bibitem[{\citenamefont{Han et~al.}(2009)\citenamefont{Han, Lai, Zi, Zhang, and
  Chan}}]{han:123904}
\bibinfo{author}{\bibfnamefont{D.}~\bibnamefont{Han}},
  \bibinfo{author}{\bibfnamefont{Y.}~\bibnamefont{Lai}},
  \bibinfo{author}{\bibfnamefont{J.}~\bibnamefont{Zi}},
  \bibinfo{author}{\bibfnamefont{Z.-Q.} \bibnamefont{Zhang}}, \bibnamefont{and}
  \bibinfo{author}{\bibfnamefont{C.~T.} \bibnamefont{Chan}},
  \bibinfo{journal}{Phys. Rev. Lett.} \textbf{\bibinfo{volume}{102}},
  \bibinfo{eid}{123904} (\bibinfo{year}{2009}).

\end{thebibliography}

\end{document}